\documentclass[aps,onecolumn,amsmath,amssymb]{revtex4}
\usepackage{graphicx}
\usepackage{dcolumn}
\usepackage{bm}

\begin{document}

\title{Nuclear shell-model code for massive parallel computation, ``KSHELL''}
\author{Noritaka Shimizu}
\email{shimizu@cns.s.u-tokyo.ac.jp}

\affiliation{Center for Nuclear Study, University of Tokyo, 
  Hongo, Tokyo 113-0033, Japan }

\begin{abstract}
A new code for nuclear shell-model calculations, ``KSHELL'', 
is developed.
It aims at carrying out both massively parallel computation
and single-node computation in the same manner.
We solve the Schr\"{o}dinger's equation in the 
$M$-scheme shell-model model space, 
utilizing Thick-Restart Lanczos method.
During the Lanczos iteration, 
the whole Hamiltonian matrix elements are generated ``on-the-fly'' 
in every matrix-vector multiplication. 
The vectors of the Lanczos 
method are distributed and stored on memory of each parallel node.
We report that the newly developed code has high parallel efficiency 
on FX10 supercomputer and a PC with multi-cores.
\end{abstract}

\maketitle

The nuclear shell-model calculations play an important role 
to understand nuclear structure \cite{ppnp_brown, review_caurier}. 
In the last two decades, 
roughly a dozen of shell-model codes had been developed
(e.g. \cite{oxbash, nushell, antoine, mshell64, bigstick, MFDn}). 
However, there is no code which is available to the public and 
applicable to MPI parallel computation still now.
I learned a lot from the code ``MSHELL64'' \cite{mshell64}, 
and developed a new code ``KSHELL'' fully 
from scratch in 2013 for massive parallel computation 
with OpenMP-MPI hybrid.
This code is written in Fortran 95 with ISO TR15581 extension.
It is equipped with dialogue-type user interface
written in python 2.

Its formalism is based on the $M$-scheme and similar to some 
preceding codes such as ``Antoine'' \cite{antoine}, ``MSHELL64'' \cite{mshell64}, 
and ``BIGSTICK'' \cite{bigstick}.
Namely, the wave function is expressed as a sum of products of 
proton and neutron configurations, 
each of which is written as an $M$-scheme state.
The matrix-vector product is performed 
with ``on-the-fly'' generation of the Hamiltonian
matrix elements. An $M$-scheme basis state is expressed 
and manipulated by 
the bit representation efficiently. The details of 
the algorithm of the $M$-scheme shell-model 
calculations are found at Ref.\cite{bigstick}.

The eigenvalue problem in the $M$-scheme basis is solved 
utilizing Thick-Restart Lanczos method \cite{tr-lanczos}, 
which is a variant of the Lanczos method and suppresses the number of the 
Lanczos vectors causing the reduction of reorthogonalization process. 
During the Lanczos iteration, the state vector is
projected out to a good total-angular-momentum state optionally. 
This projection is performed by calling the Lanczos method recursively 
to minimize the expectation value of total angular momentum.
This option is convenient for the computation of highly excited non-yrast states.

In order to fit massive parallel computation 
and to apply arbitrary truncation scheme, 
the whole $M$-scheme space is split into subspaces
by specifying the occupation numbers
of each single-particle orbit and $z$-component
of angular momentum of protons.
A unit of this subspace is called a ``partition'' or a ``sector''.
We split the matrix-vector product into 
small parts corresponding to these partitions and calculate them 
in parallel.

Its parallel performance has been tested and shows good strong scaling 
up to 8192 cores at FX10 supercomputer at the University of Tokyo.
Utilizing 8192 cores, it takes 145 seconds to compute 
the ground-state energy of $^{56}$Ni in $pf$-shell,
corresponding the eigenvalue problem of 1,087,455,228-dimension matrix.
This code copes with $O(10^{10})$ $M$-scheme dimension 
and, hopefully, larger dimension.
This code runs also on a PC 
and its performance scales almost perfectly to the number of CPU cores 
with OpenMP threads. 
Hyper-threading technology contributes little to the acceleration.

A major restriction of the large-scale 
$M$-scheme calculations is the necessity of 
large memory.
In practical calculations, 
at least two Lanczos vectors
should be stored on main memory for efficient matrix-vector multiplication.
The size of the two Lanczos vectors 
reaches huge, e.g. 80GB in the case of $10^{10}$ dimension, 
and surpasses typical memory size for a PC. 
This restriction is overcome 
by splitting the Lanczos vectors in units of ``partition'' 
and distributing them on the main memories of MPI nodes on an equal footing.
Moreover, the large number of nodes allows us to store the whole Lanczos vectors 
($\sim 100$  vectors) on memory.
It shortens the time of reorthogonalization drastically.

This code can also compute quadrupole and magnetic moments, and $E2, M1, E1$ 
transition probabilities without additional manipulation.
Three-body force is out of its purpose.

This developments are supported by a Grant-in-Aid 
for Scientific Research (25870168) from JSPS, 
the HPCI Strategic Program from MEXT, and the CNS-RIKEN joint project 
for large-scale nuclear structure calculations. 
I acknowledge Y. Utsuno (JAEA) for his contribution 
of some codes and fruitful discussions.
I also thank T. Mizusaki (Senshu Univ.) 
and M. Honma (Aizu Univ.)
for their valuable discussions and encouragements.


\end{document}